\documentstyle[psfig]{aa}
\newcommand{\ark}{\hbox{Ark~564} }

\newcommand{\etal}{et al. }
\newcommand{\asca}{{\it ASCA }}
\newcommand{\xte}{{\it RXTE }}
\def\simlt{\lower.5ex\hbox{\ltsima}}            
\def\simgt{\lower.5ex\hbox{\gtsima}}            

\def\flux{erg\,cm$^{-2}$\,s$^{-1}$}
\begin{document}
\title{Detection of a high frequency break in the X-ray power 
spectrum of Ark~564}
\author{I.E. Papadakis\inst{1,2} \and W. Brinkmann\inst{3} 
\and  H. Negoro \inst{4} \and M. Gliozzi\inst{3} }
\offprints{I. E. Papadakis;  e-mail: jhep@physics.uoc.gr}
\institute{IESL, Foundation for Research and Technology-Hellas, P.O.Box
1527, 711 10 Heraklion, Crete, Greece
\and Physics Department, University of Crete, P.O. Box 2208,
   710 03 Heraklion, Greece 
\and Max--Planck--Institut f\"ur extraterrestrische Physik,
   Giessenbachstrasse, D-85740 Garching, Germany
\and Cosmic Radiation Laboratory, RIKEN, 2-1 Hirosawa, Wako-shi,
   Saitama 351-01, Japan} 
\date{Received ?; accepted ?} 
\abstract{ 
We present a power spectrum analysis of the long \asca observation of \ark
in June/July 2001. The observed power spectrum covers a frequency range of
$\sim 3.5$ decades. We detect a high frequency break at $\sim 2\times
10^{-3}$ Hz. The power spectrum has an rms of $\sim 30\%$ and a slope of
$\sim -1$ and $\sim -2$ below and above the break frequency. When combined
with the results from a long \xte observation (Pounds \etal 2001), the
observed power spectra of \ark and Cyg X-1 (in the low/hard state) are
almost identical, showing a similar shape and rms amplitude. However, the
ratio of the high frequency breaks is very small ($\sim 10^{3-4}$),
implying that these characteristic frequencies are not indicative of the
black hole mass.  This result supports the idea of a small black hole
mass/high accretion rate in \ark.
}

\titlerunning{The X-ray power spectrum of \ark}
\authorrunning{Papadakis \etal}
\maketitle
   
\section{Introduction}
\smallskip

Narrow-line Seyfert~1 (NLS1) galaxies are a peculiar group of active
galactic nuclei (AGN) characterized by their distinct optical line
properties (Osterbrock \& Pogge 1985). In hard X-ray studies NLS1 galaxies
comprise less than 10\% of the Seyfert galaxies, however, from the ROSAT
All-Sky Survey it became clear that about half of the AGN in soft X-ray
selected samples are NLS1 galaxies (Grupe 1996, Hasinger 1997). Boller
\etal (1996) found from ROSAT observations that the soft X-ray spectra of
NLS1 galaxies are systematically steeper than those of broad line
Seyfert~1 galaxies. They further discovered that NLS1 galaxies frequently
show rapid short time scale X-ray variability which can be interpreted as
evidence for a small black hole masses in these objects.

Ark~564 is the X-ray brightest NLS1 galaxy with a 2$-$10~keV flux of $\sim
2\times10^{-11}$ \flux (Vaughan \etal 1999) and shows large amplitude
variations on short time scales (Leighly 1999). Therefore it is the best
candidate to study its X-ray variability in order to obtain important
clues about the size of the black hole mass and the accretion rate in NLS1
galaxies.  Recently, \ark was observed by \xte once every $\sim 4$ days
from January 1999 to September 2000 covering $\sim 20$ months of
data. Its $2-10$ keV power spectral density (PSD) function showed a
cut--off at a frequency which corresponds to a time scale of $\sim 13$
days (Pounds \etal 2001). \ark was also observed for a period of $\sim 35$
days in June/July 2000 by \asca. This observation was part of a
multi-wavelength AGN Watch monitoring campaign (Turner \etal 2001). The
flux variations that \ark exhibited during this observation have already
been studied by Edelson \etal (2001) who found that the variability
amplitude is almost independent of energy band and the power spectrum is
harder in the hardest energy bands. There were no delays between the
variations in the different bands and no signs of non-linear behaviour in
the light curves. Gliozzi \etal (2001) also found no sign of non-linear
behaviour and no statistically significant indication of non-stationarity
in the light curves. Furthermore, using nonlinear techniques they were
able to demonstrate that the source behaves differently in the high and
low flux states.

In this work we present a timing analysis of the June-July \asca
observation of \ark based on power spectral analysis techniques. In the
following section we present the data and the results from the PSD
analysis while in Section 3 we discuss our conclusions.

\section{Data Analysis}

\begin{figure}
\psfig{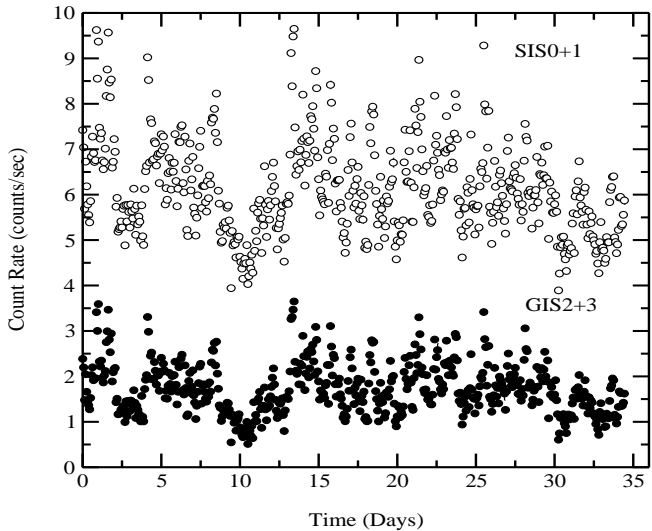}

\caption[]{ Plot of the ASCA SIS0+1 and GIS2+3 0.7--5 keV light curves
(open and filled circles, respectively). The data are binned in 5400 sec
intervals ($\sim$ the ASCA orbital period). For clarity, the SIS light
curve is shifted by $+3$ counts/sec along the $y-$axis.}

\end{figure}

ASCA observed \ark from 2000 June 01 11:51:27 to 2000 July 05 23:57:54
(UT). We extracted the X-ray data from the public archive at ISAS and
applied standard criteria for the data analysis in order to create light
curves for all four instruments. Details of the data reduction process are
given in Brinkmann \etal (2001).

Light curves were extracted in the energy bands 0.7--5 keV, 0.7--2 keV
(soft band), 2--5 keV (medium band) and 5--10 keV (hard band) for both
instruments. In all cases we combined the data from the SIS and GIS
detector pairs. Figure 1 shows the SIS0+1 and GIS2+3 0.7--5 keV light
curves in 5400 s bins ($\sim$ the \asca orbital period). Clearly, the
source is significantly variable. The root mean square variability
amplitude (corrected for the experimental noise) is $33\%, 32\%$ and
$29\%$ for the soft, medium and hard band light curves, respectively.

\subsection{Power Spectrum Analysis}

We used the 0.7--5 keV GIS and SIS light curves (which have the highest
signal to noise ratio) to estimate the overall X-ray PSD of \ark.  First,
we used the 16 sec binned SIS and GIS light curves to compute the high
frequency part of the PSD (i.e. the PSD at frequencies $> 10^{-3.5}$ Hz).  
There are 476 and 531 segments in the SIS and GIS light curves,
respectively, with no gaps in them. Their duration is between $\sim 1000$
and $\sim$ 3500 sec. For each part we computed the periodogram and
normalised it to the square of the mean count rate. Then, we combined all
the periodograms in one file, we sorted them in order of increasing
frequency, we computed their logarithm and grouped them into bins of size
$500$ following the method of Papadakis \& Lawrence (1993). To compute the
PSD at low frequencies (i.e. $<10^{-4}$ Hz) we used the 5400 sec binned
light curves. The use of the approximate \asca orbital period as the bin
size results in an evenly sampled light curve with very few missing points
(10 and 8 out of 552 in the GIS and SIS light curves, respectively). We
accounted for them using linear interpolation between the bins adjacent
to the gaps. We computed the periodograms, and averaged their logarithm
using a bin size of 20.

\begin{figure}
\psfig{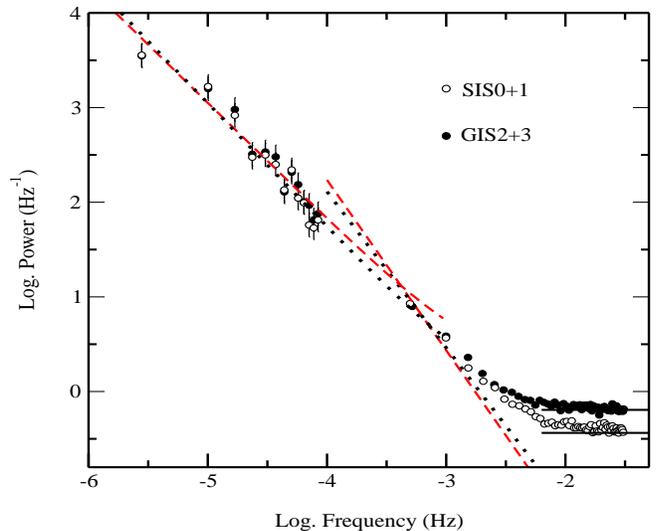} 

\caption[]{Plot of the 0.7--5 keV SIS0+1 and GIS2+3 power spectra (open
and filled circles respectively). The dotted and dashed lines show the
best fitting power law model lines to the SIS and GIS PSDs, respectively.
The solid lines show the expected Poisson noise power level for the high
frequency PSDs.}

\end{figure}

The low and high frequency PSDs of the SIS and GIS light curves are shown
in Figure 2. The SIS and GIS power spectra look similar.  Using standard
$\chi^{2}$ statistics we fitted a power law model of the form $\log
[P(\nu)]=\log[A\nu^{-\alpha}+C_{n}]$, separately to the high and low
frequency part of the PSDs (we denote with $\alpha_{\rm lf}$ and
$\alpha_{\rm hf}$ the slope of the low and high frequency parts
respectively). The constant $C_{n}$ represents the contribution of the
experimental Poisson noise process to the observed PSD. It can be
estimated using the average error of the points in each light curve (e.g.
Papadakis \& Lawrence 1995) and, as a result, $C_{n}$ was kept fixed at
its expected value (shown with solid lines in Figure 2 for the high
frequency power spectra)  during the model fitting procedure. The best
fitting results are listed in Table 1 and the best fitting models are
plotted in Figure 2 (dotted and dashed lines for the SIS and GIS power
spectra, respectively). All errors quoted in this paper correspond to the
$68\%$ confidence region and were computed following the method of Lampton
et al. (1976).

The best fitting slope values of the SIS and GIS power spectra are
consistent within their errors ($\alpha^{\rm SIS}_{\rm hf}-\alpha^{\rm
GIS}_{\rm hf}=-0.15\pm 0.07$ and $\alpha^{\rm SIS}_{\rm lf} - \alpha^{\rm
GIS}_{\rm lf}=0.08\pm 0.18$). Therefore, despite the slight difference in
the responses of the detectors on board \asca, the power spectra of the
GIS and SIS light curves are indeed similar. The weighted mean
$\alpha_{\rm hf}$ and $\alpha_{\rm lf}$ values are $1.70\pm0.03$ and
$1.27\pm0.09$, respectively. Their difference, $0.43\pm0.10$, shows that
the low and high frequency parts of the power spectra have significantly
different slopes. As Figure 2 shows, the slope of the PSD changes at $\sim
10^{-3}$ Hz. Above that frequency, the slope is significantly steeper than
the slope at lower frequencies.

\begin{table}
\begin{center}
\caption{Results of the power law model fits to the 0.7--5 keV, SIS and
GIS power spectra of \ark. }
\begin{tabular}{lcccc} \hline
 & $\alpha_{\rm lf}$ & $\alpha_{\rm hf}$ & $\chi^{2}_{\rm lf}$/dof
&$\chi^{2}_{\rm hf}$/dof \\
\hline
SIS0+1 & $1.31\pm0.13$ & $1.65\pm0.04$ & 15.6/11 & 78/57 \\
GIS2+3 & $1.23\pm0.13$ & $1.80\pm0.06$ & 13.2/11 & 106/59 \\
\hline
\end{tabular}
\end{center}
\end{table}

Since the SIS and GIS power spectra are similar, we combined them to
produce the final, 0.7--5 keV PSD of \ark. The resulting power spectrum
(with the Poisson noise subtracted) is shown in Figure 3. We have grouped
the combined GIS and SIS periodograms into bins of size 200, 600 and 1000
for the high frequency part of the spectrum (the bin size is increased
towards the higher frequencies in order to increase the signal to noise)
and into bins of size 50 for the low frequency part. A simple power law
model cannot fit the PSD well ($\chi^{2}=259$ for 58 dof). For that
reason, we used a ``broken power law" model. According to this model, the
low frequency part of the PSD is fit to a function of the form $A
\nu^{-\alpha_{\rm low}}$ and, above a ``break frequency" $\nu_{\rm bf}$,
the PSD is fitted to a power law with a different index, $\alpha_{\rm
high}$. The fit results are listed in Table 2; the best fit model is
plotted in Figure 3 (solid line). Compared to the simple power law model,
a broken power law provides a significantly better fit to the data. The
PSD of \ark has a $\sim -1.2$ slope at frequencies up to $\sim 2\times
10^{-3}$ Hz above which the PSD steepens to a slope of $\sim -2$.

Finally, we investigated whether the PSD parameters depend on energy.  We
used the 0.7--2 keV, 2--5 keV and 5-10 keV light curves to compute the
power spectra in the three bands. A power law model cannot fit the soft
and medium band PSDs ($\chi^{2}=206$ and 157 for 66 and 53 dof,
respectively). The broken power law model gives a significantly better fit
to these power spectra. The hard band PSD cannot be computed for
frequencies higher than a few $\times 10^{-3}$ Hz due to the low signal to
noise of the $16$-sec hard band light curves; therefore the simple power
law model gives already an acceptable fit to the PSD in this case. The
best fitting results for the power law (for the hard band PSD) and broken
power law models (for the soft and medium band PSDs) are listed in Table
2. The model parameters of the soft and medium energy PSDs are consistent
within the errors, except from $\alpha_{\rm low}$. Our results show that
the low frequency PSDs become significantly harder with increasing energy.
This is in agreement with the results of Edelson \etal (2001).

\begin{figure}
\psfig{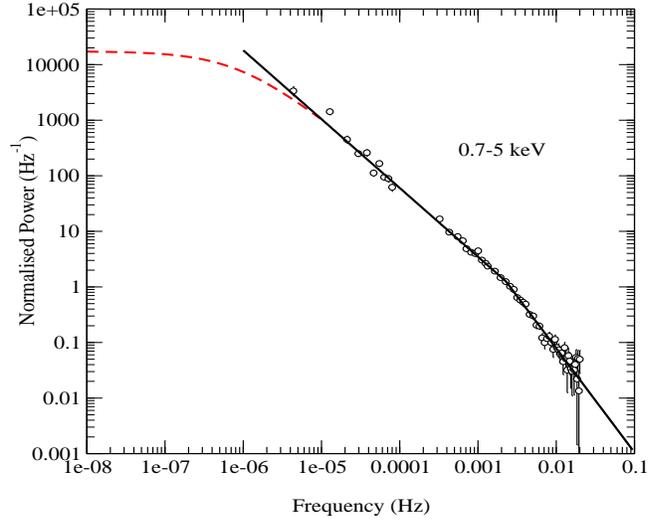}

\caption[]{Plot of the combined GIS and SIS 0.7--5 keV power spectra. The
solid line shows the best fitting ``broken power law" model to the data.
The dashed line is the best fitting model to the \xte PSD (from Pounds
\etal 2001).}

\end{figure}

\begin{table}
\begin{center}
\caption{Results of the ``broken power law" model fits to the PSDs of
\ark}
\begin{tabular}{lccccc}
\hline
Energy  & $\alpha_{\rm low}$ & $\alpha_{\rm high}$ & 
$\nu_{\rm bf}$ & $\chi^{2}$/dof \\
 & & & $(\times 10^{-3}$ Hz) & \\
\hline
0.7--5 keV  & $1.24^{+0.03}_{-0.04}$ & $1.90^{+0.17}_{-0.14}$ 
& $2.3^{+0.6}_{-0.6}$ &  72.6/56 \\
0.7--2 keV  & $1.26^{+0.03}_{-0.04}$ & $1.98^{+0.24}_{-0.20}$ 
& $2.6^{+0.6}_{-0.6}$ & 84.1/64 \\
2--5 keV  & $1.18^{+0.03}_{-0.04}$ & $1.87^{+0.34}_{-0.32}$ 
& $1.7^{+0.6}_{-0.9}$ & 84.1/51 \\
5--10 keV  & $1.13^{+0.06}_{-0.03}$ & -- & -- & 52.9/31 \\
\hline
\end{tabular}
\end{center}
\end{table}

\section{Discussion}

Using the \asca June-July 2001 observation of \ark we have estimated its
power spectrum over a frequency range of $\sim 3.5$ decades. We have
detected a high frequency break at $\sim 2\times 10^{-3}$ Hz.  

In Figure 3, together with the $0.7-5$ keV PSD, we have also plotted the
best fitting ``cut-off power law" model from Pounds \etal (2001).  The
observed PSD of \ark extends over $\sim 6$ decades in frequency. It is
clear from this Figure that the PSD starts to flatten below $\sim 10^{-5}$
Hz, showing a second, low frequency break.  MCG 6-30-15 also shows two
break frequencies in its PSD (Nowak \& Chiang 2000), however, the evidence
for the low frequency break is not significant in that case (Uttley \etal
2001). Therefore, \ark is probably the only Seyfert galaxy for which we
are certain that the observed PSD has a shape that is almost identical to
the shape of the Cyg X-1 PSD in the low/hard state. In both cases, the PSD
slope is flat below a characteristic low frequency. Above this frequency
the PSD steepens to $\propto \nu^{-1}$ and then becomes $\propto \nu^{-2}$
above a second break frequency.  Apart from the PSD shape, the rms
amplitudes at frequencies higher than the low frequency cut off are also
very similar. These results show that \ark corresponds to the low/hard
(instead of the high/soft) state of the Galactic black hole candidates
(GBHC).

However, despite this similarity, \ark is not simply a ``larger" version
of GBHCs.  The two break frequencies in Cyg X-1 are located between $0.03-
0.3$ and $1-10$ Hz (e.g. Nowak \etal 1999, Negoro \etal 2001). The ratio
between the characteristic frequencies in \ark and Cyg X-1 is therefore
$\sim 10^{5-6}$ and $\sim 10^{3-4}$, if we consider the low and high
frequency breaks, respectively.  Although there is an uncertainty
associated with these ratios (for example, we do not know whether the
break frequencies in AGN are constant or change with time) this cannot
probably explain the two orders of magnitude difference between the
ratios. Perhaps then the break frequencies do not scale proportionally
with the black hole mass. For example, if the high frequency break in Cyg
X-1 corresponds to the Keplerian orbital period at $\sim 30 R_{S}$ around
the central object (Nowak \etal 1999; $R_{S}$ is the Schwarzschild radius)
and the high frequency break in \ark corresponds to the same time scale
but at a radius $\sim 3 R_{S}$, then the ratio of the characteristic time
scales will not be equal to the mass ratio but to the mass ratio $\times
(30 R_{S}/3 R_{S})^{-3/2}$.

On the other hand, if the low frequency breaks scale with mass, then the
central black hole mass of \ark should be $\sim 10^{7}$ M$_{\odot}$,
assuming that the black hole in Cyg X-1 is $10$ M$_{\odot}$ (Herrero et
al. 1995). If that is the case, taking into account the luminosity of
\ark, Pounds \etal (2001) concluded that the source is accreting at a
substantial fraction of the Eddington limit. The orbital period at
$3R_{S}$ for a $10^{7}$ M$_{\odot}$ black hole is $\sim 5000$ sec, which
is an order of magnitude larger than the time scale which corresponds to
the high frequency break in the PSD of \ark. Either the mass of the black
hole in \ark is smaller than $10^{7}$ M$_{\odot}$ (in which case the
accretion rate is larger than the Eddington limit), or due to the large
accretion rate, the characteristic time scales are smaller. When the
accretion rate approaches the Eddington limit advective energy transport
dominates over radiative cooling and the disk becomes moderately
geometrically thick, a so called ``slim disk" (Abramowicz et al. 1988). In
this case, a substantial amount of radiation can be produced from inside
the last stable orbit, i.e. $3R_{S}$ (Mineshige et al. 2000). The
free-fall time scale at distance $R$ is given by $t_{\rm
ff}=(R_{G}/c)(R/R_{G})^{3/2}=500$M$_{8}(R/R_{G})^{3/2}$ sec (Rees 1984),
where M$_{8}$ is the mass in units of $10^{8}$M$_{\odot}$ and
$R_{G}=GM/c^{2}$ is the gravitational radius. For M$_{8}=0.1$ and
$R=3R_{S}=6R_{G}$, we get $t_{\rm ff}\sim 700$ sec which is comparable to
$1/\nu_{\rm bf}$ that we find for \ark. The sound crossing time scale is
another possible candidate that could correspond to $1/\nu_{\rm bf}$. In
the thin disk case this time scale is very long even in the innermost
parts of the disk. However, when the accretion rate is comparable to the
Eddington rate and the disk is dominated by radiation pressure, the group
velocity of sound waves (responsible for transmitting density fluctuation
information) becomes very large (e.g. Krolik et al. 1991). Consequently,
the sound crossing time scale at say $R=3R_{S}$ will be decreased
substantially and could be of the order of $1/\nu_{\rm bf}$.

We conclude that, the similarity of the PSD and time lags in \ark implies
that X-rays in this source are produced by a mechanism similar to the X-ray
emission mechanism in Cyg X-1 in its low/hard state. On the other hand, the
comparison between the characteristic time scales in the two systems
suggests that at least one physical parameter is different in them,
probably the accretion rate, with \ark having a substantially higher
accretion rate than Cyg X-1 in its low/hard state.

\vskip 0.4cm
\begin{acknowledgements}
Part of this work was done in the TMR research network
'Accretion onto black holes, compact stars and protostars' funded by the
European Commission under contract number ERBFMRX-CT98-0195.  
\end{acknowledgements}


\begin{thebibliography}{}
\bibitem{} Abramowicz, M.A., Czerny, B., Lasota, J.P., Szuszkiewicz,
E. 1988, ApJ, 332, 646
\bibitem{} Boller, T., Brandt, W.N., Fink, H. 1996, A\&A, 305, 53
\bibitem{} Brinkmann, W., Papadakis, I.E., Negoro, H., Detsis, E.,
Papamastorakis, I., Gliozzi, M., Scheingraber, H. 2001,
in Proc. Maxi Workshop on AGN variability, in press
\bibitem{} Gliozzi, M., Brinkmann, W., R\"ath C., Papadakis, I.E., Negoro,
H., Scheingraber, H. 2001, A\&A, submitted
\bibitem{} Edelson, R., Turner, T.J., Pounds, K., Vaughan, S., Markowitz,
A., Marshall, H., Dobbie, P., Warwick, R. 2001, ApJ, in press
\bibitem{} Grupe, D. 1996, PhD thesis, Univ. G\"ottingen
\bibitem{} Hasinger, G. 1997, in Imaging and Spectroscopy of Cosmic Hot
Plasmas, ed. F. Makino, \& K. Mitsuda (UAP, Tokyo), 263
\bibitem{} Herrero, A., Kudritzki, R.P., Gabler, R., Vilchez, J.M., Gabler,
A. 1995, A\&A, 297, 556
\bibitem{} Krolik, J.H., Horne, K., Kallman, T.R., Malkan, M.A., Edelson,
R.A., Kriss, G. A. 1991, ApJ, 371, 541
\bibitem{} Lampton, M., Margon, B., Bowyer, S. 1976, ApJ, 208, 177 
\bibitem{} Leighly, K.M. 1999, ApJS, 125, 297
\bibitem{} Mineshige, S., Kawaguchi, T., Takeuchi, M., Hayashida, K. 2000,
PASJ, 52, 499
\bibitem{} Negoro, H., Kitamoto, S., Mineshige, S. 2001, ApJ, 554, 528
\bibitem{} Nowak, M.A., Vaughan, B.A., Wilms, J., Dove, J.B., Begelman,
M. 1999, ApJ, 510, 874
\bibitem{} Nowak, M.A., Chiang, J. 2000, ApJ, 531, L13     
\bibitem{} Osterbrock, D.E., Pogge, R.W. 1985, ApJ, 297, 166
\bibitem{} Papadakis, I.E, Lawrence, A. 1993, MNRAS, 261, 612 
\bibitem{} Papadakis, I.E., Lawrence, A. 1995, MNRAS, 272, 161
\bibitem{} Pounds, K., Edelson, R., Markowitz, A., Vaughan, S. 2001, ApJ
550, L15 
\bibitem{} Rees, M.J. 1984, ARA\&A, 22, 471 
\bibitem{} Turner, T.J., Romano, P., George, I.M., Edelson, R., Collier,
S.J., Mathur, S., Peterson, B.M. 2001, ApJ, 561, 131
\bibitem{} Uttley, P., McHardy, I.M., Papadakis, I.E. 2001, MNRAS,
submitted
\bibitem{} Vaughan, S., Reeves, J., Warwick, R., Edelson, R. 1999, MNRAS
309, 113
\end{thebibliography}
\end{document}